\documentclass[aps,pra,twocolumn,superscriptaddress,showpacs,floatfix]{revtex4}
\usepackage{graphicx,dcolumn,bm,hyperref,amsmath,amssymb,xspace,epsfig,float,array,multirow,amsfonts,mathrsfs}
\usepackage{color}
\usepackage{amstext}
\usepackage{latexsym}

\usepackage{braket} 
\usepackage{ulem}
\usepackage{bbm}

\begin{document}
\title{\bf Tan's contact scaling behaviour for trapped Lieb-Liniger bosons: from two to many\\
}

\author{Matteo Rizzi}
\affiliation{Johannes Gutenberg-Universit\"at, Institut f\"ur Physik, Staudingerweg 7, 55099 Mainz, Germany}

\author{Christian Miniatura}
\affiliation{
 Universit\'e C\^ote d'Azur, CNRS, Institut de Physique de Nice, 1361 route des Lucioles, 06560 Valbonne, France
}
\affiliation{MajuLab, CNRS-UCA-SU-NUS-NTU International Joint Research Unit, Singapore
}
\affiliation{
Centre for Quantum Technologies, National University of Singapore, 117543 Singapore, Singapore
}
\affiliation{
Department of Physics, National University of Singapore, 2 Science
Drive 3, Singapore 117542, Singapore}
\affiliation{
School of Physical and Mathematical Sciences, Nanyang Technological University, 637371 Singapore, Singapore
}

\author{Anna Minguzzi}
\affiliation{Univ. Grenoble-Alpes, CNRS, LPMMC, 38000 Grenoble, France}

\author{Patrizia Vignolo}
\affiliation{
 Universit\'e C\^ote d'Azur, CNRS, Institut de Physique de Nice, 1361 route des Lucioles, 06560 Valbonne, France
}

\date{\today}

\begin{abstract}
  We show that the contact parameter of $N$ harmonically-trapped interacting 1D 
bosons at zero temperature can be analytically and accurately obtained
  by a simple rescaling of the exact two-boson solution, and that $N$-body
  effects can be almost factorized. 
  The small deviations observed between our analytical results and DMRG
  calculations are more pronounced when the interaction energy is maximal
  (i.e. at intermediate interaction strengths) but they remain bounded by
  the large-$N$ local-density approximation obtained from the Lieb-Liniger
  equation of state stemming from the Bethe Ansatz.
  The rescaled two-body solution is so close to the exact ones, that
  is possible, within a simple expression
  interpolating the rescaled two-boson result to the local-density, to obtain
   $N$-boson contact and ground state energy functions in very good
  agreement with DMRG calculations.
  Our results suggest a change of
  paradigm in the study of interacting quantum systems, giving
  to the contact parameter a more fundamental role than energy.  
\end{abstract}
\maketitle
 
\section{Introduction}
The relation between two-body and many-body physics is often an important point for the comprehension and the description of strongly-correlated quantum systems.
A celebrated example is provided by homogeneous one-dimensional (1D) interacting systems solvable by the Bethe Ansatz, such as bosons and fermions
with contact interactions~\cite{LiebLin,Yang67,Sutherland68}. In that case, the $N$-body solution can be exactly expressed as a function of a product of two-body scattering
contributions. Generally, such a system is no longer integrable when subjected to an external potential but a notable exception is
the limit of infinitely strong repulsive interactions, known as the Tonks-Girardeau limit, where fermionization occurs. In that case, the system remains exactly solvable, for any number of bosons and fermions
\cite{Vignolo00,Deuretzbacher,Fang2011,vignolo2013,Volosniev2014,Deuretzbacher2014,Decamp2016,Decamp2016b,Decamp2017}. 
At finite interactions, the harmonically-trapped system can be exactly solved for 2 particles~\cite{Busch98} and
is approximately solved in the
large-$N$ limit by a local density approximation (LDA) on
the Lieb-Liniger solution \cite{Olshanii03}.
For finite-$N$ systems, several approaches have been proposed: a pair-correlated
wavefunction approach \cite{Brouzos2012,Koscik2018};
a $T$-matrix approach for the Fermi polaron at zero and finite temperature
\cite{Doggen2013};
a geometric wavefunction description, that is very accurate
for 2 and 3 bosons \cite{Wilson2014}; and, more recently, 
an interpolatory Ansatz combining the non-interacting and unitary wavefunctions \cite{Andersen2016}.
This last approach provides very accurate results for the energy in 
impurity systems \cite{Andersen2016}, but is less accurate when increasing
the number of particle components \cite{Pecak2017}.

A crucial observable for a 1D system of $N$ particles with contact interactions
is Tan's contact parameter, characterizing the asymptotic behavior of the momentum distribution of the particles  $C_{N}= \lim_{k\to \infty} k^4 n(k)$~\cite{Minguzzi02}. The contact
embeds information on the interaction energy and 
the density-density correlation function \cite{Tan2008a,Tan2008b,Tan2008c}. 
It is a univocal measure of the wavefunction symmetry of fermionic and/or bosonic mixtures~\cite{Decamp2016b,Decamp2017}.
The contact parameter is also determined by the probability density 
of finding 2 particles at a vanishing distance \cite{Olshanii03}.
For trapped quantum gases, this probability density has a nontrivial dependence on the number of particles and on the interaction
strength~\cite{xu2015,Yao2018}.

In this Letter, we propose a change of paradigm 
by showing that the contact parameter plays in fact a more fundamental 
role than the energy in analyzing Lieb-Liniger bosons.
Inspired by the scaling properties of this model,
we show that if the starting point of the scaling analysis
is the contact parameter instead of the energy, the two-body result
provides a very good description of the system
for {\it any} number of particles and interaction strengths.
The quantitative difference between our predictions and numerically-exact DMRG results is always very small (i.e. less than a few percent) and 
is the largest at intermediate interaction strengths where the interaction energy is also the largest.
For particle numbers $N>2$ we show that the many-body corrections to the rescaled two-body result
can be accounted for  by a simple interpolation connecting the 
two-body solution and the LDA one.
With this, we obtain an analytical and very accurate expression for the contact parameter at all particle numbers $N$ that we use to derive an accurate formulation for the total energy of the system.

\section{Model and scaling analysis}
We start with the case of  $N \geq 2$ identical and harmonically-trapped
1D bosons of mass $m$ 
at zero temperature, interacting via repulsive contact interactions. Such a system is described by the many-body Hamiltonian 
\begin{equation}
H=\sum_{j=1}^N \left[ \frac{-\hbar^2}{2m}\frac{\partial^2}{\partial x_j^2} + \frac{1}{2} m \, \omega^2 \, x_j^2 
	+g \sum_{\ell>j} \delta(x_j-x_\ell)\right]
 \label{eq:Ham}
\end{equation}
with $g = 2\hbar^2/(m |a_{\textrm{1D}}|) \geq 0$~\cite{Olsh98}. As shown by Tan in~\cite{Tan2008a,Tan2008b,Tan2008c}, the contact parameter associated to the eigenenergy $E_{N}$ reads
\begin{equation}
 C_{N}(g)=	\frac{m^2}{\pi\hbar^4} \left( - \frac{\partial E_{N}}{\partial g^{-1}}\right)  = 
  			\frac{m^2 g^2}{\pi\hbar^4}  \frac{\partial E_{N}}{\partial g} \equiv
			\frac{m^2 g}{\pi\hbar^4}  E_\textrm{int} \, 
\label{eq:contact}
\end{equation} 
where $E_\textrm{int} $ is the interaction energy. Tan's contact Eq.\eqref{eq:contact} is thus a direct by-product of the dependence of the system energy on the interaction strength $g$.
In the following, we will stick our analysis to the ground state energy. 
By rescaling Hamiltonian (\ref{eq:Ham}) by the ground state energy
in the fermionized regime $E^\infty_{N} = N^2\hbar\omega/2$, and by expressing the particle coordinates
in units of $a_{\textrm{ho}}/\sqrt{N}$, where $a_{\textrm{ho}}=\sqrt{\hbar/(m\omega)}$ is the harmonic oscillator
length, it is easy to see that the ground state energy writes~\cite{Decamp2016b} 
\begin{equation}
E_{N}(g)=E^\infty_{N} \, \mathcal{E}(N,g_N)
\label{eq:einf}
\end{equation}
where 
\begin{equation}
g_N = \dfrac{mga_\textrm{ho}}{2 \hbar^2 \sqrt{N}} = \dfrac{a_{\textrm{ho}}}{|a_{\textrm{1D}}|\sqrt{N}} \equiv \dfrac{\alpha}{\sqrt{N}}
\label{eq:AlphaN}
\end{equation} 
is the dimensionless interaction strength and $\alpha\!=\!a_{\textrm{ho}}/|a_{\textrm{1D}}|$. 
The dimensionless energy function $\mathcal{E}$ interpolates between the non-interacting regime where $\mathcal{E}(N,0)=1/N$ and the fermionized regime where $\mathcal{E}(N,\infty)=1$. Obviously, Tan's contact depends on the same parameters $N$ and $g_N$ and reads:
\begin{equation}
 C_{N}(g)=
 \dfrac{N^{5/2}}{\pi a_{\textrm{ho}}^3} \, \mathcal{C}(N,g_N),
 \label{eq:reducedTan0}
\end{equation}
where the rescaled dimensionless Tan's contact  
\begin{equation}
  \mathcal{C}(N,z) = z^2 \, \partial_z \mathcal{E}(N,z)
\label{eq:reducedTan}
\end{equation}
is evaluated at $z=g_N$.
By the same token, $E_\textrm{int} = E^\infty_{N} \, \mathcal{E}_\textrm{int}(N, z)$ and we find
\begin{equation}
\mathcal{E}_\textrm{int}(N, z) = \dfrac{\mathcal{C}(N, z)}{z} = z \, \partial_z \mathcal{E}(N,z).
\label{eq:InterEn}
\end{equation}

In the thermodynamic limit ($N, a_\textrm{ho} \to \infty$ at constant $a_\textrm{ho}/\sqrt{N}$), the only scaling parameter
is $g_N$, both for the dimensionless energies $\mathcal{E}$, $\mathcal{E}_\textrm{int}$ and contact parameter $\mathcal{C}$.
This can be easily shown in a Local Density Approximation (LDA) on  the Lieb-Liniger homogeneous solution \cite{LiebLin,Olshanii03},
and generalized to a generic trapping potential (see App.~\ref{app-scaling}).
One gets:
\begin{equation}
\begin{split}
  & E^{\textrm{\tiny LDA}}_{N}(g)=E^\infty_{N} \, \mathcal{E}_{\textrm{\tiny LDA}}(g_N) \\
& C^{\textrm{\tiny LDA}}_{N}(g)=\dfrac{N^{5/2}}{\pi a_{\textrm{ho}}^3} \, \mathcal{C}_{\textrm{\tiny LDA}}(g_N)
  \label{eq:scalingC}
  \end{split}
\end{equation}
with 
$\mathcal{E}_{\textrm{\tiny LDA}}(0)\! = \! \mathcal{C}_{\textrm{\tiny LDA}}(0)\! = \!0$, $\mathcal{E}_{\textrm{\tiny LDA}}(\infty)\! = \!1$ and $\mathcal{C}_{\textrm{\tiny LDA}}(\infty)\! = \!128\sqrt{2}/(45\pi^2)$~\cite{Olshanii03}.
Although the derivation has been detailed for single-component  bosons, it is possible to show that the scaling analysis applies also to multi-component bosons and fermions \cite{Massignan2015,Matveeva2016,Decamp2016b,Lewenstein-Massignan,Laird2017}, the Hamiltonian being the same as Eq. (\ref{eq:Ham}).

\section{The reduced contact parameter and scaling Ansatz}
\begin{figure*}
\begin{center}
  \includegraphics[width=0.45\linewidth]{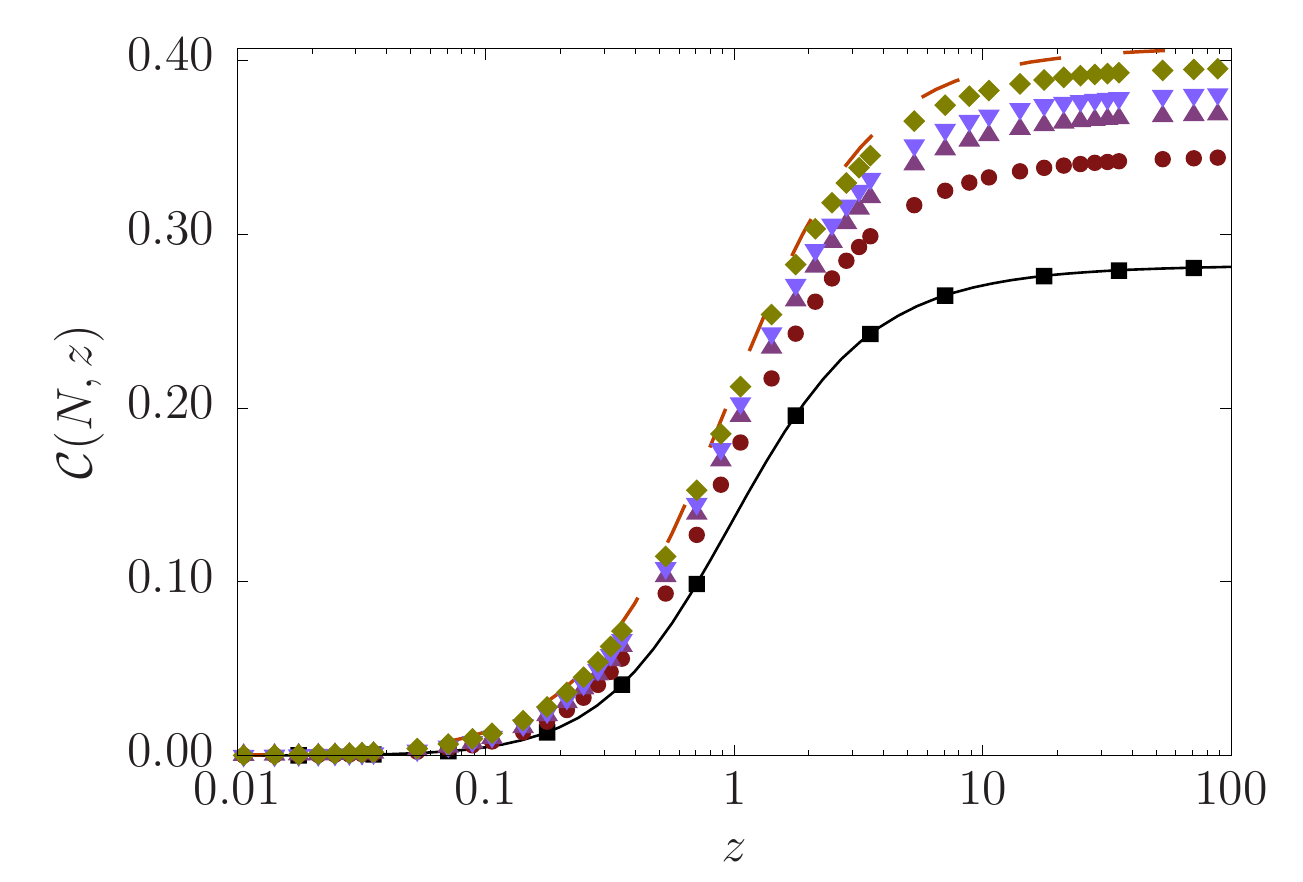}
 \includegraphics[width=0.45\linewidth]{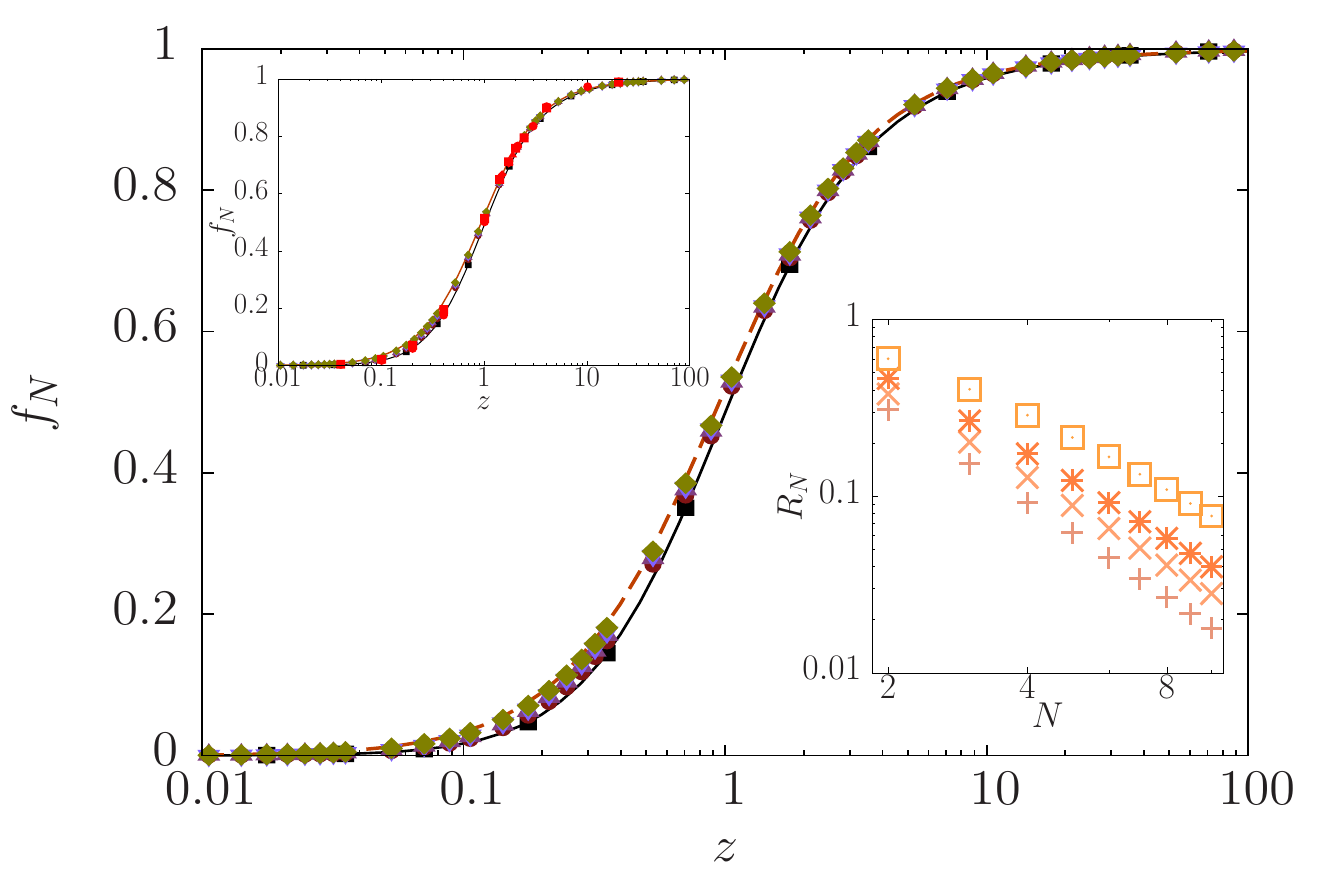}
\end{center}
\caption{\label{fig1} (Color online) Rescaled dimensionless contact $\mathcal{C}(N,z)$, Eq.\eqref{eq:reducedTan}, (left panel) and reduced contact parameter $f_{N}(z)$, Eq.\eqref{eq:scalingHyp}, (right panel) as a function of the dimensionless scaling parameter $z=a_{\textrm{ho}}/(|a_{\textrm{1D}}|\sqrt{N})=a_{\textrm{ho}}/(|a_{\textrm{1D}}'|\sqrt{2})$.
The different symbols corresponds to DMRG calculations: 
$N=2$ (black squares), $N=3$ (brown circles),
  $N=4$ (purple triangles up), $N=5$ (light-blue triangles down) and
  $N=8$ (green diamonds).
  The black continuous line corresponds to Eq.~\eqref{eq:c2}, and
  the orange dashed line corresponds to the LDA solution, Eq.~\eqref{eq:scalingC}~\cite{Olshanii03}.
  Top inset in the right panel: Reduced contact parameter $f_{N}(z)$ for $SU(\kappa)$
  fermions \cite{Decamp2016b} (red points), $\kappa$ ranging from 2 to 6,
  superposed to all the data and
  curves of the main panel.
  Bottom inset in the right panel:
 Convergence rate  $R_{N}(z)$ as a function of $N$, in a log-log scale, for $z=$ 0.14 (squares), 0.35 (stars), 0.70 (crosses), and $z\rightarrow\infty$ (plus).}
\end{figure*}

Strictly speaking, the LDA scaling behavior with respect to the sole variable
$z$ should only hold in the large-$N$ limit.
Indeed, it is what we observe if we plot $\mathcal{C}(N,z)$ obtained by a 2-tensor DMRG optimisation of a Matrix Product States (MPS) 
Ansatz ~\cite{Schollwock2011} (see App.~\ref{app-dmrg}) in comparison with $\mathcal{C}_{\rm LDA}(z)$, as shown in the left panel of Fig. \ref{fig1}.
However all the curves seem to
have the same shape, but with different asymptotic values. Here we put
forward a different {\it scaling hypothesis} by assuming that
the reduced scaling parameter
\begin{equation}
  f_{N}(z)=\dfrac{C_N(g(z))}{C_N(\infty)}=\dfrac{\mathcal{C}(N,z)}{\mathcal{C}(N,\infty)},
\end{equation}
with $g(z)=2\hbar^2\sqrt{N}z/(ma_{\textrm{ho}})$,
is an universal function for any $N\ge 2$.
In particular, if this scaling hypothesis holds,
\begin{equation}
f_{N}(z) = f_2(z).
      \label{eq:scalingHyp}
\end{equation}
This would correspond to the assumption that a $N$-boson system at contact interaction strength $g$ is amenable to an effective 2-boson system at a rescaled {\it weaker} contact interaction strength $g'=\sqrt{2/N} g$.
Stated equivalently, the scattering length is renormalized through $a_{{\textrm 1D}} \to a_{{\textrm 1D}}'=\sqrt{N/2} \, a_{{\textrm 1D}}$.
In the case of $N=2$ bosons, Tan's contact is given by 
\begin{equation}
  C_2(g)=\dfrac{m^2g^2}{\pi\hbar^4} |\psi_\nu(0)|^2
  \label{eq:c2}
\end{equation}
where
$\psi_\nu(0)$
is the wavefunction solving the Schr\"odinger equation for the relative motion~\cite{Busch98}
evaluated at $x_1-x_2=0$. 
It is straightforward (see App. \ref{app-two}) to show that 
\begin{equation}
   f_2(z)=\dfrac{\mathcal{C}(2,z)}{\mathcal{C}(2,\infty)} =
	\dfrac{\pi \nu^2\, 2^{\nu-1}}{\mathcal{N}(\nu) \, [\Gamma(1-\nu/2)]^2} 
\label{scaling1}
\end{equation}
where $\mathcal{C}(2,\infty)= 1/(2\sqrt{\pi})$, $\mathcal{N}(\nu)$ is a normalization factor (see App. \ref{app-two}), and the $\nu$'s solve
\begin{equation}
\dfrac{\Gamma(-\nu/2)}{\Gamma(-\nu/2+1/2)} =-\dfrac{1}{z}.
\label{gammaeq}
\end{equation}

In the right panel of Fig. \ref{fig1}, we compare the exact result for $f_2(z)$, Eq.~\eqref{scaling1}, to the numerical data.
The fact that all curves (almost) collapse show that
  $z=a_{\textrm{ho}}/(|a_{\textrm{1D}}|\sqrt{N})=a_{\textrm{ho}}/(|a_{\textrm{1D}}'|\sqrt{2})$
is indeed the dimensionless scaling parameter of the reduced contact parameter, and that the contact for any interaction strength and any number $N$ of particles can be deduced from a simple 2-body calculation, $f_2(z)$, {\it and} from the knowledge of the contact for $N$ particles in the Tonks-Girardeau limit, $\mathcal{C}(N,\infty)$, that, for bosons, can be calculated exactly \cite{vignolo2013}.
This means also that the function $C_N(\infty)$ almost embeds
the full $N$-dependence of the problem for any value of $z$, even for few-body systems where the $N^{5/2}$ factor, deduced in the thermodynamic limit,
starting from the energy scaling-analysis, fails.
This result seems to be general and not to depend on the particle statistics \cite{Massignan2015,Matveeva2016,Decamp2016b,Lewenstein-Massignan,Laird2017}.
Indeed, the data for the reduced contact
parameter of a harmonically-trapped one-dimensional SU($\kappa$)
interacting fermions \cite{Decamp2016b} collapse on the same curve,
as shown in the top inset in the right panel of Fig. \ref{fig1}.

\subsection{Are two enough?}
Our DMRG data match at first sight very well with the simple prediction of Eq.~\eqref{eq:scalingHyp}. 
    However, we observe small deviations at intermediate interaction strengths where the data lie between 
    $f_2$ (black continuous line) and the LDA solution
    $f_{\textrm{\tiny LDA}}=\mathcal{C}_{\textrm{\tiny LDA}}(z)/\mathcal{C}_{\textrm{\tiny LDA}}(\infty)$
    (orange dashed line) that is known to be
    a very good approximation for the contact in the large-$N$ limit. 
    This point is illustrated in the bottom inset
    of the right panel of Fig. \ref{fig1},
    where we show, by plotting the convergence rate  $R_{N}(z) =  1- {\mathcal{C}(N,z)}/{\mathcal{C}_{\textrm{\tiny LDA}}(z)}$, how fast the exact contact converges to its LDA value at increasing  $N$, for various values of $z$.
    A numerical fit in the fermionized regime \cite{vignolo2013} gives

    \begin{equation}
     R_{N}(\infty) =  1- \dfrac{\mathcal{C}(N,\infty)}{\mathcal{C}_{\textrm{\tiny LDA}}(\infty)} \simeq  1.04 \, N^{-7/4}.
     \label{eq:conv}
    \end{equation}
    The weak dependence on $z$ of the slope of the convergence rate $R_{N}(z)$
    confirms that the dependence on $N$ of $C_N(z)$ is almost
    independent of $z$.
    
  \subsection{Beyond two}
\begin{figure}
  \begin{center}
   \includegraphics[width=1\linewidth]{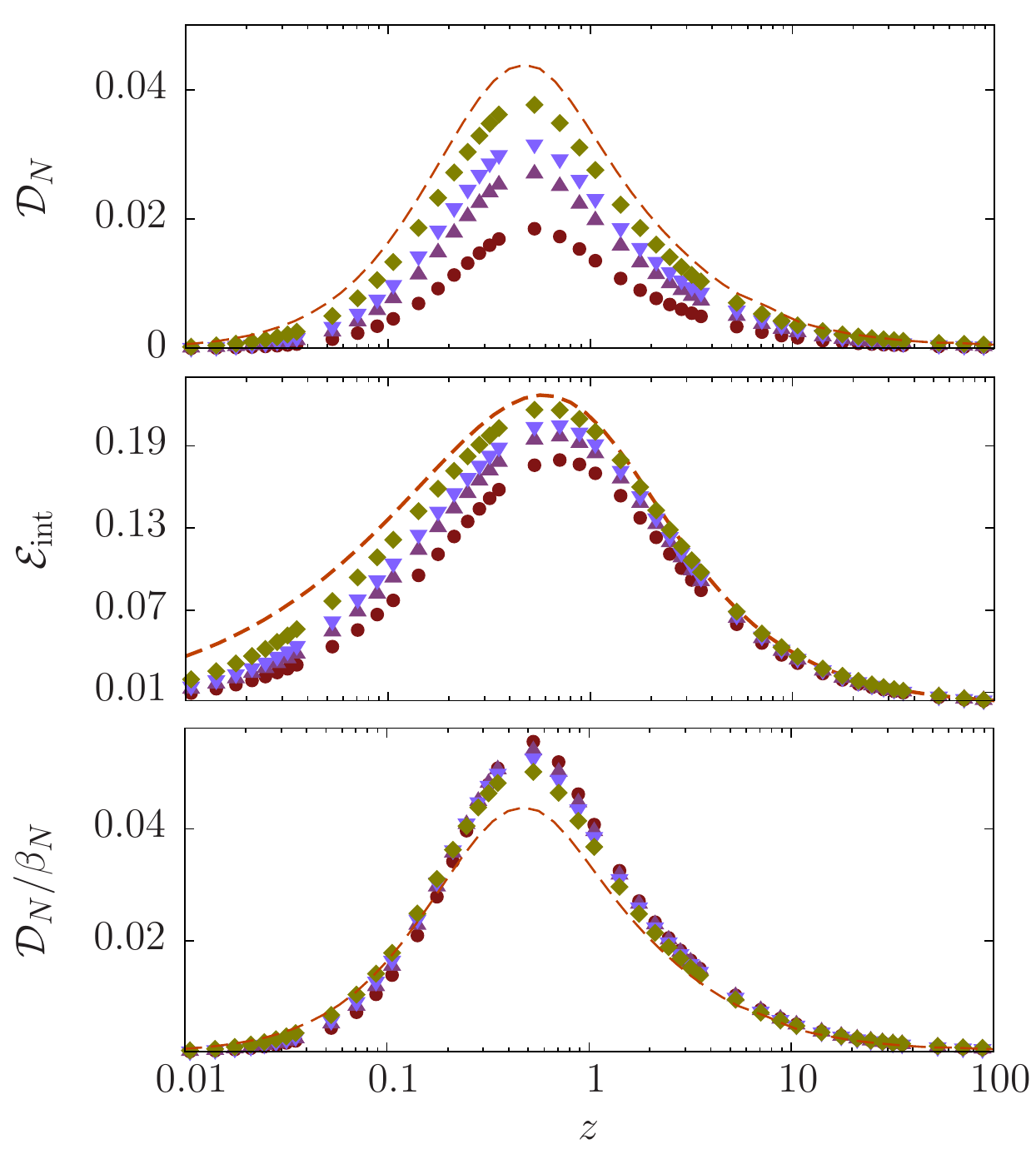}
    \end{center}
  \caption{\label{fig2}
Top panel: (Color online) Difference $\mathcal{D}_{N}(z) \!=\! f_{N}(z) \!-\! f_2(z)$ for different values of $N$ as a funciton of the dimensionless scaling parameter $z$. 
Middle panel: Dimensionless interaction energy $\mathcal{E}_{\textrm{int}}(N,z)$ for different values of $N$, Eq.\eqref{eq:InterEn}, as a function of $z$. All curves display a clear maximum at intermediate dimensionless interaction strengths $z \simeq 0.5$.
Bottom panel: Scaled difference $\mathcal{D}_{N}(z)/\beta_N$ with $\beta_N=1-2/N$ for different values of $N$ as a function of $z$. All curves collapse quite well onto the LDA prediction $\mathcal{D}_{\textrm{\tiny LDA}}(z)$
(dashed orange curve) even if further corrections would be needed around the
maximum.
Symbols are the same as in Fig.~\ref{fig1}.}
\end{figure}

\begin{figure}
\begin{center}
  \includegraphics[width=1\linewidth]{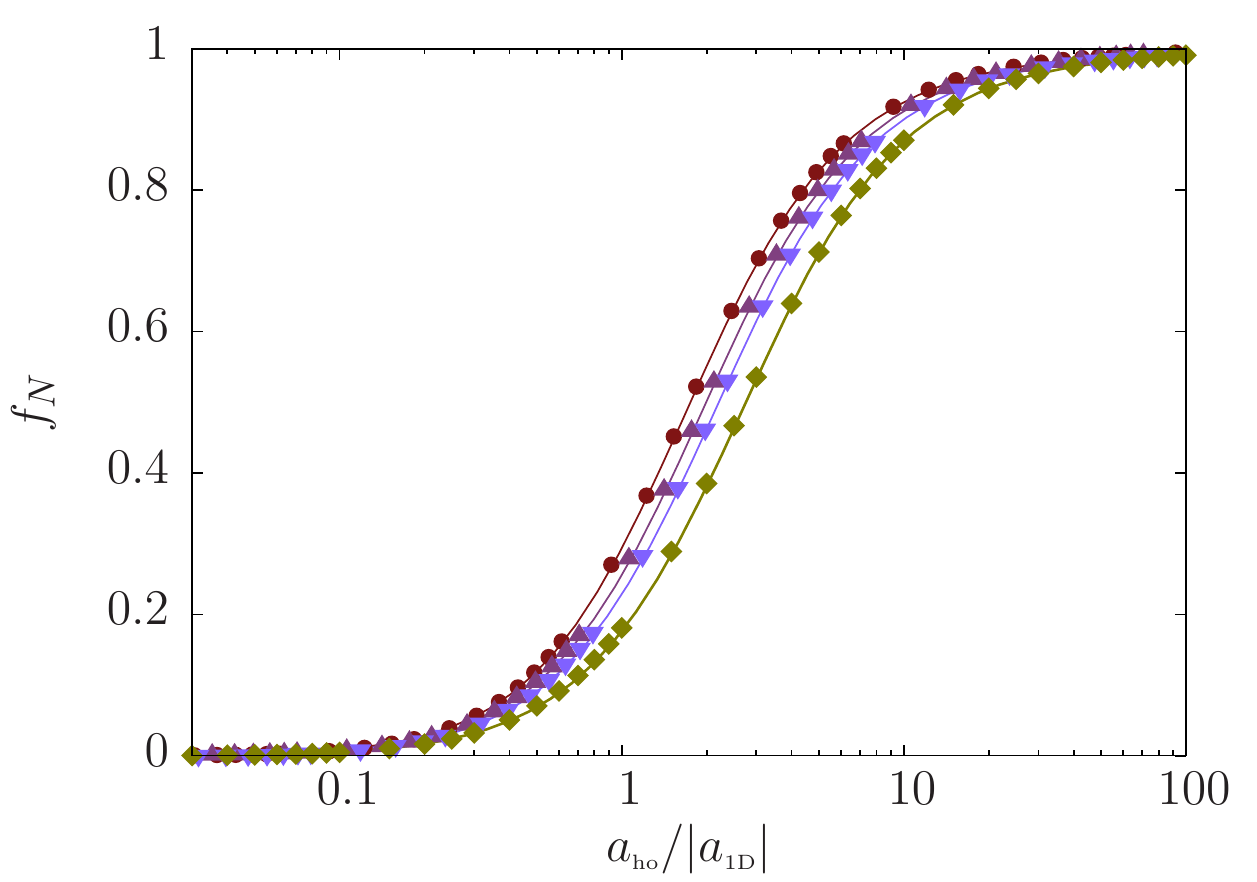}
    \end{center}
\caption{\label{fig3} (Color online) Reduced contact parameter $f_{N}(\alpha/\sqrt{N})$, Eq.\eqref{eq:scalingHyp}, for different $N$ and plotted as a function of $\alpha\!=\!a_{\textrm{ho}}/|a_{\textrm{1D}}|$ for better visibility. 
  Solid lines: theoretical prediction Eq.~\eqref{2corr}; symbols: DMRG results. Symbols are the same as in Fig.~\ref{fig1}.
}
    \end{figure}

  To further quantify
  the corrections to the scaling prediction Eq.\eqref{eq:scalingHyp}, 
  we plot in Fig. \ref{fig2} the difference 
  $\mathcal{D}_{N}(z) \!=\! f_{N}(z) \!-\! f_2(z)$.
We observe that $\mathcal{D}_{N}(z)$ reaches its largest value where the
interaction energy $\mathcal{E}_{\textrm{int}}(N,z)$ is maximum. By comparing $\mathcal{D}_{N}(z)$ to the LDA prediction $\mathcal{D}_{\textrm{\tiny LDA}}(z) = f_{\textrm{\tiny LDA}}(z) - f_2(z)$ (orange dashed line), 
we infer the approximate, but quite accurate, proportionality relation $\mathcal{D}_{N}(z) \simeq \beta_N \, \mathcal{D}_{\textrm{\tiny LDA}}(z)$ with $\beta_N = 1-2/N$, see bottom panel of Fig. \ref{fig2}. 
As a consequence, the simple interpolation 
\begin{equation}
f_N(z) \simeq \left(1-\beta_N\right) \, f_2(z)
  +\beta_N\, f_{\textrm{\tiny LDA}}(z)
 \label{2corr}
\end{equation}
connects quite accurately the exact two-body solution for the contact parameter to the LDA one. 
We validate this interpolation in  Fig. \ref{fig3} by comparing Eq.~\eqref{2corr} with DMRG data obtained for $N=3, 4, 5$ and $8$ bosons. We find a
perfect agreement. This means that, within our approach, we can calculate with
the same degree of precision all non-trivial experimentally relevant quantities
that are directly connected
to the contact parameter, such as the interaction energy \cite{Tan2008a,Zwe11},
the two-body correlation function \cite{Olshanii03,Zwe11},
the magnetization \cite{Decamp2016b}, the loss-rate in boson-fermion
mixtures \cite{Sebastien2017}, or the heating rate due to measurement
back-action of an atomic system in an optical cavity \cite{Uchino2018}.

    \section{From the contact to the energy} 
\begin{figure}
\begin{center}
  \includegraphics[width=1\linewidth]{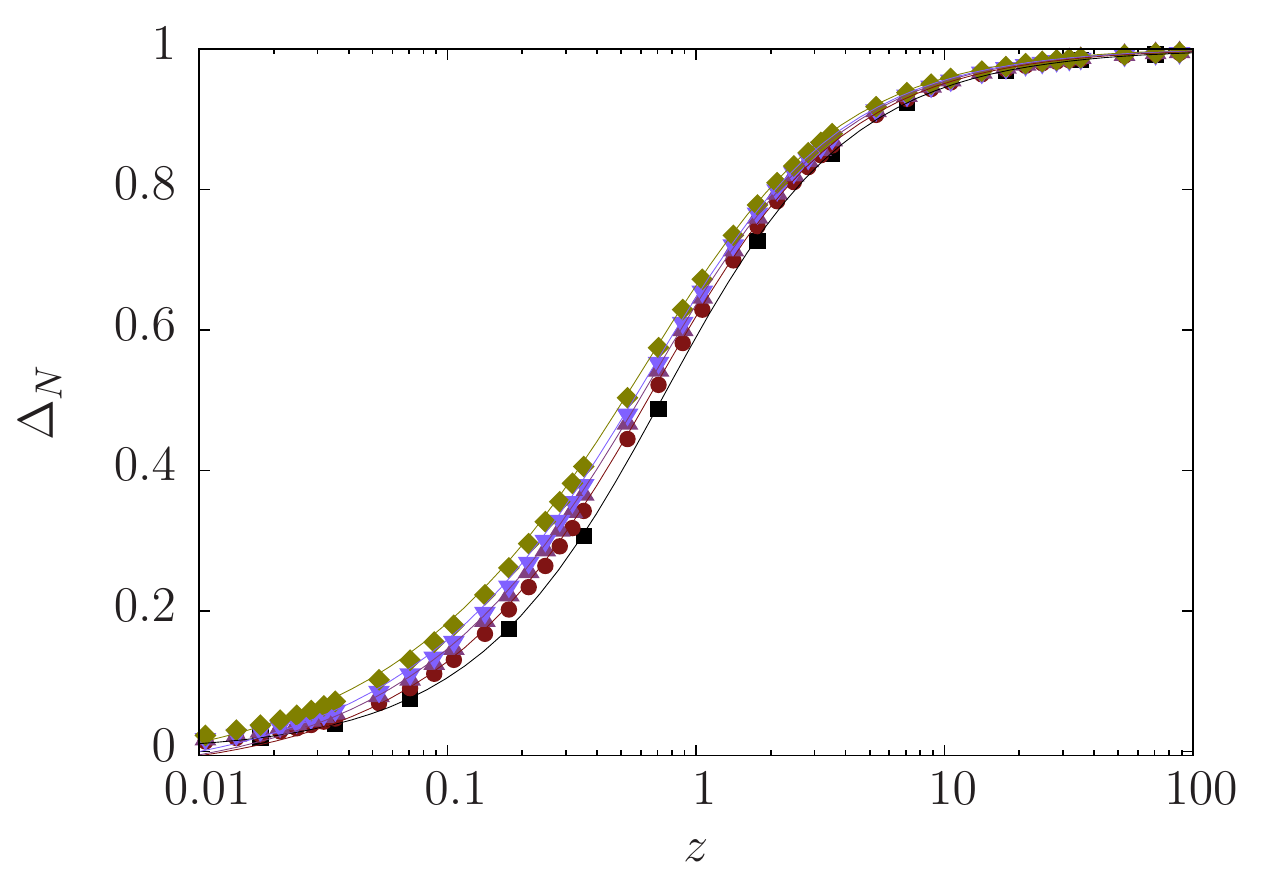}
\end{center}
\caption{\label{fig4} (Color online) Rescaled ground-state energy  $\Delta_N(z)$ relative to its non-interacting  value, Eq.(\ref{miaDiff}), as a function of the dimensionless scaling parameter $z$  for different values of $N$.
  Solid lines: theoretical prediction Eq.~\eqref{mia}; symbols: DMRG results, same $N$ values  and symbols   as  in Fig.~\ref{fig1}. 
}
\end{figure}

    The most crucial test of the quality of our Ansatz for the contact
    parameter is the ground-state energy, since it is obtained by integration
    of the contact adding up the deviations:
\begin{equation}
\mathcal{E}(N, z) = 1 - \int_{z}^\infty dz' \, \dfrac{\mathcal{C}(N,z')}{z'^2}.
\label{eq-en}
\end{equation}
Using Eq.~\eqref{2corr}, we arrive at 
\begin{equation}
\begin{split}
 \mathcal{E}(N,z) &\simeq 1 - \dfrac{2}{N} \, \dfrac{\mathcal{C}(N,\infty)}{\mathcal{C}(2,\infty)}\left[1- \mathcal{E}(2, z)\right]\\
&-\left(1-\dfrac{2}{N}\right) \, \dfrac{\mathcal{C}(N,\infty)}{\mathcal{C}_{\textrm{\tiny LDA}}(\infty)} \left[1- \mathcal{E}_{\textrm{\tiny LDA}}(z)\right].
\end{split}
\label{mia}
\end{equation}
In Fig. \ref{fig4}, we plot the rescaled  energy difference
\begin{equation}
    \Delta_N(z) =\dfrac{E_N(g(z))-E_N(0)}{E_N^\infty-E_N(0)}=
   \dfrac{N \mathcal{E}(N, z) - 1}{N-1},
\label{miaDiff}
\end{equation} 
whose limits $\Delta_N(\infty) \!=\! 1$ and $\Delta_N(0) \!=\! 0$ do not depend on $N$.
We compare the exact numerical results with
the prediction obtained by using Eq.\eqref{mia} for different values of $N$ .
The agreement with the DMRG data is very good from moderately weak to strong interaction strengths ($z \ge 0.02$).  
Discrepancies only occur in the weak interaction regime ($z \le 0.02$) where LDA is less accurate.

\section{Conclusion}
  We have shown that the contact parameter for $N$ harmonically-trapped interacting 1D bosons at zero temperature can be simply
  and accurately obtained from an appropriate rescaling of the two-body contact parameter followed by a smooth interpolation to the $N$-body LDA one. The key point is a change of paradigm: identifying the contact as the starting point 
for the scaling analysis instead of the energy. 
Indeed almost all the dependence of the contact
  on the number of particles can be embedded
  in the contact at infinite interactions for {\it any} number of particles. 
This result seems to be general and not to depend on the particle statistics. 
It shows the fundamental role
of the contact, that is likely due to its
 local two-body correlation nature. 
  We have further shown that our approach leads to a ground state energy for any number of bosons that matches very well the exact result  down to moderately weak interaction strengths where no analytical solution is known. 
Our results improve on previous studies~\cite{Brouzos2012,Wilson2014,Andersen2016,Pecak2017} with a simpler and more accurate Ansatz, that further confirm that the ground state properties of an interacting 1D Bose gas
can be accurately described by an effective two-body contact interaction dressed by the other particles in the fluid \cite{Braaten2008,Zwe11}.
Our work constitutes an important step forward in understanding the effects 
of correlations and interactions in harmonically-trapped
one-dimensional interacting boson and fermion mixtures.
It opens the way to further studies of similar scaling 
properties in higher-dimensional systems \cite{Levinsen2017}, confined in various trapping potentials, at zero and finite temperature \cite{Yao2018}.
  
  
  \section*{Acknowledgements}
  P.V. acknowledges UMI 3654 MajuLab hospitality and D. Goupy for enlightening discussions. A.M. aknowledges ANR SuperRing project  (ANR-15-CE30-0012-02), and discussions with G. Lang.
  M.R. acknowledges computational time from the Mogon cluster of the JGU (made available by the CSM and AHRP),
  S. Montangero for a long-standing collaboration on the flexible Abelian Symmetric Tensor Networks Library employed here, as well as J. J\"unemann for his participation in early stages of this work. C.M. is a Fellow of the Institute of Advanced Studies at Nanyang Technological University (Singapore). The Centre for Quantum Technologies is a Research Centre of Excellence funded by the Ministry of Education and National Research Foundation of Singapore.


  \appendix
  \section{Scaling properties and Local Density Approximation for Tan's contact parameter}
\label{app-scaling}

We detail here the derivation of the scaling properties of $N$ one-dimensional
bosons with contact interactions of strength $g$. \\

\subsection{Scaling for the homogeneous system} For a homogeneous system of length $L$, the number density $\rho = L/N$ defines a length scale $\rho^{-1}$ and an energy scale $\hbar^2\rho^2/(2m)$. Scaling all spatial variables by $\rho^{-1}$ in the Hamiltonian Eq.(\ref{eq:Ham}) with $\omega =0$, it is easy to see that both the energy per particle and the energy density follow, in the thermodynamic limit, the scaling relations
\begin{equation}
\dfrac{E_N}{N} = \dfrac{\hbar^2\rho^2}{2m} \, e(\gamma), \qquad \dfrac{E_N}{L} = \dfrac{\hbar^2\rho^3}{2m} \, e(\gamma),
\label{eq:ScalingRel}
\end{equation}
where $e(\gamma)$ is a monotonically increasing function of the dimensionless interaction strength 
\begin{equation}
\gamma=\dfrac{mg}{2\hbar^2 \rho}=\dfrac{1}{\rho |a_{\mathrm{1D}}|} =  \dfrac{L}{N |a_{\mathrm{1D}}|} \equiv \dfrac{\alpha}{N}
\end{equation}
where $\alpha = L/|a_{\mathrm{1D}}|$.
The scaling relations Eq.\eqref{eq:ScalingRel} and the equation of state $e(\gamma)$ for the homogeneous system were exactly determined by Lieb and Liniger via the Bethe Ansatz~\cite{LiebLin}. In the thermodynamic limit $N, L \to \infty$ at constant density $\rho$, it takes values  between $e(0)=0$ and $e(\infty) = \pi^2/3$. 
For $g\to\infty$, and for large $N$, we have $E^\infty_N \simeq N^3 \epsilon_0 /3$ where $\epsilon_0 = \pi^2\hbar^2/(2mL^2)$ is the ground-state energy of a particle in a box of size $L$. Then, from Tan's relation for the contact parameter, see Eq.(\ref{eq:contact}), it is easy to infer: 
\begin{equation}
  C_N(g) \simeq \dfrac{N\rho^3}{\pi} \, \gamma^2 \dfrac{de}{d\gamma} = \dfrac{N^4}{\pi L^3} \, \gamma^2 \dfrac{de}{d\gamma}
\label{eq:contactapp}
\end{equation}
for the homogeneous system. Note that, following Eq.(\ref{eq:einf}), we would have $\mathcal{E}(N, \gamma) \simeq  3e(\gamma)/\pi^2$ for the homogeneous system at large $N$. \\

\subsection{Scaling for the harmonically-trapped system}
In the presence of a harmonic potential, the appropriate thermodynamic limit is instead obtained by taking $N \to \infty$ and $\omega \to 0$ at constant $N\epsilon_0$ where $\epsilon_0=\hbar\omega/2$ is now the harmonic groundstate energy~\cite{Decamp2016b}. Stated equivalently, $N$ and $a_\mathrm{ho}\to\infty$ at constant ratio $\sqrt{N}/a_\mathrm{ho}$. Note that this ratio can be interpreted as an effective (constant) particle density $\rho= N/L_N$ in the thermodynamic limit for a system of size $L_N = \sqrt{N}a_\mathrm{ho} \to \infty$. Using this $\rho$ and $\hbar^2\rho^2/(2m)$ as the spatial and energy scales of the system, considerations analogous to the homogeneous case then lead to Eqs.(\ref{eq:einf}-\ref{eq:AlphaN}) with $E^\infty_{N} = N^2\hbar\omega/2$, $\alpha = a_\mathrm{ho}/|a_{\mathrm{1D}}|$ and $\gamma = \alpha/\sqrt{N}$ $(\equiv g_N)$. In particular, Eq.\eqref{eq:contactapp} immediately leads to Eq.(\ref{eq:reducedTan0}-\ref{eq:reducedTan}) when replacing $\rho$ by $\sqrt{N}/a_\mathrm{ho}$.

Our approach is an alternative to the one developed in \cite{xu2015} where the scaling is expressed as a function of the parameter $\gamma(0)=mg/(\hbar^2 \rho(0))$ where  $\rho(0)$ is the density at the trap center. \\

\subsection{Scaling for a general trapping potential}
Let us considering the case of an arbitrary confining potential $V(x)$,
in the case where the wavefunction vanishes at the boundaries.
Denoting by $\epsilon_n = \eta_n \, \epsilon_\xi$ ($n\in \mathbbm{N}$) the consecutive energy levels of $V(x)$, where $\epsilon_\xi= \hbar^2/(2m\xi^2)$ and $\xi$ are the characteristic energy and length scales of the trap, the thermodynamic limit is obtained in a similar way. Indeed, the ground state energy per particle in the infinitely-repulsive interacting limit then reads $E_N(g=\infty)/N = b(N) \,  \epsilon_\xi$ where 
\begin{equation}
b(N) = \dfrac{1}{N} \ \sum_{n=0}^{N-1} \eta_n \sim N^{2q}.
\end{equation}
The thermodynamic limit is then obtained by taking $N\to\infty$ and $\xi\to\infty$ at constant ratio $N^{q}/\xi$. We would thus have $L_N = \xi \, N^{1-q}$ and $\gamma \equiv g_N = \alpha \, N^{-q}$. For the harmonic trap, one has $q=1/2$.\\

\subsection{Local density approximation (LDA)}
Such scaling forms for the harmonically-trapped system are recovered exactly in the LDA.
We start from the chemical potential of the homogeneous system as obtained from the Lieb-Liniger equation of state:
\begin{equation}
\mu_\mathrm{h}=\dfrac{\partial E_N}{\partial N} = \frac{\hbar^2\rho^2}{2m} \, \left( 3 \, e(\gamma) - \gamma  \frac{de}{d\gamma} \right).
\label{eq:mu}
\end{equation}
By defining the interaction energy scale $\epsilon_g=\hbar^2/(2m a_{\mathrm{1D}}^2) = 4g/|a_{\mathrm{1D}}|$, we see that $\mu_\mathrm{h} = \epsilon_g \, F(\gamma)$ with:
\begin{equation}
F(\gamma) = 3 \dfrac{e(\gamma)}{\gamma^2} - \dfrac{1}{\gamma} \dfrac{de}{d\gamma}. 
\end{equation}
The above is a monotonous function of $\gamma$ for bosons in the Lieb-Liniger model~\cite{LiebLin}. Inverting this equation, we can obtain the particle density in terms of the chemical potential under the form {$\rho |a_{\mathrm{1D}}|= n(\mu_\mathrm{h}/\epsilon_g)$}, where $n$ is a dimensionless function.
In the presence of the harmonic potential $V_{\mathrm{ext}}(x)= m\omega^2x^2/2$, the inhomogeneous density profile within the LDA reads
\begin{equation}
\rho(x) = \frac{1}{|a_{\mathrm{1D}}|} \ n \left( \frac{\mu_\mathrm{t} - V_{\mathrm{ext}}(x)}{\epsilon_g} \right) \Theta(1- |x|/R),
\end{equation} 
where $\Theta$ is the Heaviside step function and $R=\sqrt{2 \mu_t/(m\omega^2)}$ the Thomas-Fermi radius.

The chemical potential of the trapped gas $\mu_\mathrm{t}$ is obtained by imposing the normalization condition
$N = \int \rho(x) \mathrm{d}x$. After the change of variable $z=x/R$, and noting that $\epsilon_g =  \hbar\omega \, \alpha^2/2$ and $R= a_\mathrm{ho} \, \sqrt{2\mu_t/(\hbar\omega)}$ it is easy to recast this normalization condition into 
\begin{equation}
 \sqrt{\dfrac{\mu_t}{\epsilon_g}} \, 
 	\int_{|z|\leq 1} n\left[ \dfrac{\mu_t}{\epsilon_g} (1-z^2) \right] \, \mathrm{d}z  = \dfrac{1}{g^2_N}.
\label{eq:scalingMU}
\end{equation}
Just like for the homogeneous case, this equation can be inverted to give $\mu_t = \epsilon_g \, M(g_N)$.
By integrating backwards the chemical potential, $E_N(g) \equiv \int_0^N \mu_\mathrm{t}(N') \, \mathrm{d}N'$, the dimensionless LDA energy writes
\begin{equation}
	\mathcal{E}_{\mathrm{LDA}}(g_N) = \dfrac{2 E_N(g)}{\hbar\omega N^2} = \dfrac{\alpha^2}{2} \, \int_0^N M\left(\alpha/\sqrt{N'}\right) \,  dN'. 
\end{equation}
With the change of variables $y=\alpha/\sqrt{N'}$, we finally arrive at
\begin{equation}
	\mathcal{E}_{\mathrm{LDA}}(g_N) = 2 g_N^4 \, \int_{g_N}^\infty \dfrac{M(y)}{y^3} dy,
\label{eq:scalingEN}	
\end{equation}
from which the LDA Tan's contact parameter $\mathcal{C}_{\mathrm{LDA}}(z) = z^2 \, \partial_z \mathcal{E}_{\mathrm{LDA}}(z)$ follows. 
For bosons in the Tonks-Girardeau regime, one has $\mathcal{C}_{\textrm{\tiny LDA}}(\infty)\! = \!128\sqrt{2}/(45\pi^2)$~\cite{Olshanii03}. The corresponding expressions for multicomponent fermions in the limit of infinite repulsive interactions have been derived in \cite{Decamp2016b}.

\section{Density Matrix Renormalization Group (DMRG)}
\label{app-dmrg}
The numerical results for the Tan's contact of several particles at finite interactions
have been obtained by a two-tensor DMRG optimisation of a 
Matrix Product States (MPS) Ansatz~\cite{Schollwock2011}.
Namely, we take a (tight-binding) lattice discretization of Eq.(\ref{eq:Ham})
in a sufficiently large box ($L$ up to 12 $a_\mathrm{ho}$), 
and we extract the continuum limit by considering lattice spacings $a$ down to $a_\mathrm{ho}/16$:
the tunneling amplitude, external potential and on-site interaction strength scale like $t \propto a^{-2}$, $V \propto a^{2}$, and $U \propto a^{-1}$ respectively.
We encompass the conservation laws of the particle number in the tensor network structure directly, 
in order to achieve both speed-up and increased accuracy.
The discarded probability is kept below $10^{-12}$, 
and no truncation is performed on the local bosonic Hilbert space.
For more details, we refer the reader, e.g., to a recent work of ours~\cite{Decamp2016b}.

\section{Reduced contact parameter for 2 bosons}
\label{app-two}
In the case of $N=2$ bosons, Tan's contact is given by 
\begin{equation}
  C_2(g)=\dfrac{m^2g^2}{\pi\hbar^4} |\psi_\nu(0)|^2
  \label{eq:c2}
\end{equation}
where
\begin{equation}
  \psi_\nu(0)=\dfrac{(\pi/2)^{1/4}}{\sqrt{a_{\textrm{ho}}}\sqrt{\mathcal{N}(\nu)}}\dfrac{2^{\nu/2}}{\Gamma(-\nu/2+1/2)} \, 
  \Phi\left(-\dfrac{\nu}{2},\dfrac{1}{2},0\right)
\end{equation}
is the wavefunction solving the Schr\"odinger equation for the relative motion~\cite{Busch98}
evaluated at $x_1-x_2=0$. $\Gamma(u)$ is the gamma Euler function, $\Phi$ is the (Kummer) hypergeometric function, and
\begin{equation}
\mathcal{N}(\nu)= \Gamma (\nu +1)  \left\{ 1 + \tfrac{\sin (\pi  \nu)}{2\pi} 
	\left[ \mathtt{\Psi}\left(\tfrac{\nu}{2} + 1\right) - \mathtt{\Psi}\left(\tfrac{\nu}{2} + \tfrac{1}{2}\right) \right] 
	\right\}
\end{equation}%
is a normalization factor involving the digamma function $\mathtt{\Psi}(u) = \Gamma^\prime(u) / \Gamma(u)$.
It is straightforward to show that 
\begin{equation}
   \dfrac{\mathcal{C}(2,g_2)}{\mathcal{C}(2,\infty)} =
	\dfrac{\pi \nu^2\, 2^{\nu-1}}{\mathcal{N}(\nu) \, [\Gamma(1-\nu/2)]^2} \equiv f_2(g_2) \, 
\label{scaling1}
\end{equation}
where $\mathcal{C}(2,\infty)= 1/(2\sqrt{\pi})$. The $\nu$'s are indeed a
function of $\alpha/\sqrt{2}$ since they solve
\begin{equation}
\dfrac{\Gamma(-\nu/2)}{\Gamma(-\nu/2+1/2)} = \dfrac{\sqrt{2}}{\alpha} =\dfrac{1}{g_2} 
\label{gammaeq}
\end{equation}
and are the analog of the integers labeling the Hermite polynomials in the harmonic oscillator~\cite{Busch98}.
Noticeably, for $\nu \, \in \, [0,1]$, we have $\mathcal{N}(\nu) \simeq 1$, 
$[\Gamma(1-\nu/2)]^{-2}\simeq [1-(1-1/\pi)\nu]$ and
\begin{equation}
 f_2(g_2)\simeq \nu^2 \, 2^{\nu-1} \, [\pi-(\pi-1)\nu].
      \label{fappr}
\end{equation}


\begin{thebibliography}{10}

\bibitem{LiebLin}
E. Lieb and W. Liniger, Phys. Rev. {\bf 130},  1605  (1963).

\bibitem{Yang67}
C.~N. Yang, Phys. Rev. Lett. {\bf 19},  1312  (1967).

\bibitem{Sutherland68}
B. Sutherland, Phys. Rev. Lett. {\bf 20},  98  (1968).

\bibitem{Vignolo00}
P. Vignolo, A. Minguzzi, and M. Tosi, Phys. Rev. Lett. {\bf 85},  2850  (2000).

\bibitem{Deuretzbacher}
F. Deuretzbacher {\it et~al.}, Phys. Rev. Lett. {\bf 100},  160405  (2008).

\bibitem{Fang2011}
B. Fang {\it et~al.}, Phys. Rev. A {\bf 84},  023626  (2011).

\bibitem{vignolo2013}
P. Vignolo and A. Minguzzi, Phys. Rev. Lett. {\bf 110},  020403  (2013).

\bibitem{Volosniev2014}
A. Volosniev {\it et~al.}, Nature Communications {\bf 5},  5300  (2014).

\bibitem{Deuretzbacher2014}
F. Deuretzbacher {\it et~al.}, Phys. Rev. A {\bf 90},  013611  (2014).

\bibitem{Decamp2016}
J. Decamp {\it et~al.}, New Journal of Physics {\bf 18},  055011  (2016).

\bibitem{Decamp2016b}
J. Decamp {\it et~al.}, Physical Review A {\bf 94},  053614  (2016).

\bibitem{Decamp2017}
J. Decamp {\it et~al.}, New Journal of Physics {\bf 19},  125001  (2017).

\bibitem{Busch98}
T. Busch, B.-G. Englert, K. Rz\c{a}\.{z}ewski, and M. Wilkens, Found. Phys.
  {\bf 28},  549  (1998).

\bibitem{Olshanii03}
M. Olshanii and V. Dunjko, Phys. Rev. Lett. {\bf 91},  090401  (2003).

\bibitem{Brouzos2012}
I. Brouzos and P. Schmelcher, Phys. Rev. Lett. {\bf 108},  045301  (2012).

\bibitem{Koscik2018}
P. Ko\'scik, M. Plodzie\'n, and T. Sowi\'nski, arXiv:1804.06342  (2018).

\bibitem{Doggen2013}
E.~V.~H. Doggen and J.~J. Kinnunen, Phys. Rev. Lett. {\bf 111},  025302
  (2013).

\bibitem{Wilson2014}
B. Wilson {\it et~al.}, Phys. Lett. A {\bf 378},  1065  (2014).

\bibitem{Andersen2016}
M.~E.~S. Andersen {\it et~al.}, Scientific Reports {\bf 6},  28362  (2016).

\bibitem{Pecak2017}
D. P\c{e}cak, A.~S. Dehkharghani, N.~T. Zinner, and T.
  Sowi\ifmmode~\acute{n}\else \'{n}\fi{}ski, Phys. Rev. A {\bf 95},  053632
  (2017).

\bibitem{Minguzzi02}
A. Minguzzi, P. Vignolo, and M. Tosi, Phys. Lett. A {\bf 294},  222  (2002).

\bibitem{Tan2008a}
S. Tan, Ann. Phys. (N.Y.) {\bf 323},  2971  (2008).

\bibitem{Tan2008b}
S. Tan, Ann. Phys. (N.Y.) {\bf 323},  2987  (2008).

\bibitem{Tan2008c}
S. Tan, Ann. Phys. (N.Y.) {\bf 323},  2952  (2008).

\bibitem{xu2015}
W. Xu and M. Rigol, Phys. Rev. A {\bf 92},  063623  (2015).

\bibitem{Yao2018}
H. Yao {\it et~al.}, arXiv:1804.04902  (2018).

\bibitem{Olsh98}
M. Olshanii, Phys. Rev. Lett. {\bf 81},  938  (1998).

\bibitem{Massignan2015}
P. Massignan, J. Levinsen, and M.~M. Parish, Phys. Rev. Lett. {\bf 115},
  247202  (2015).

\bibitem{Matveeva2016}
N. Matveeva and G. Astrakharchik, New Journal of Physics {\bf 18},  065009
  (2016).

\bibitem{Lewenstein-Massignan}
T. Grining {\it et~al.}, Phys. Rev. A {\bf 92},  061601  (2015).

\bibitem{Laird2017}
E.~K. Laird, Z.-Y. Shi, M.~M. Parish, and J. Levinsen, Phys. Rev. A {\bf 96},
  032701  (2017).

\bibitem{Schollwock2011}
U. Schollw\"ock, Ann. Phys. {\bf 326},  96  (2011).

\bibitem{Zwe11}
M. Barth and W. Zwerger, Ann. Phys. {\bf 326},  2544  (2011).

\bibitem{Sebastien2017}
S. Laurent {\it et~al.}, Phys. Rev. Lett. {\bf 118},  103403  (2017).

\bibitem{Uchino2018}
S. Uchino, M. Ueda, and J.-P. Brantut, arXiv:1802.04024  (2018).

\bibitem{Braaten2008}
E. Braaten and L. Platter, Phys. Rev. Lett. {\bf 100},  205301  (2008).


\bibitem{Levinsen2017}
J. Levinsen, P. Massignan, S. Endo, and M.~M. Parish, Jour. of Phys. B {\bf
  50},  072001  (2017).

\end{thebibliography}

\end{document}